\documentclass[espf]{aa}

\usepackage{color} 

\usepackage{graphicx}
\usepackage{amsmath}
\usepackage[psamsfonts]{euscript}
\usepackage[psamsfonts]{amssymb}
\usepackage{amsmath}
\usepackage{graphicx}
\usepackage{natbib}
\usepackage{epsfig}

\titlerunning{}

\def\kms{km~s$^{-1}$}

\begin{document}
\title{High resolution spectroscopy for Cepheids distance
determination\thanks{Based on observations made with ESO telescopes
at the Silla Paranal Observatory under programme IDs 072.D-0419 and
073.D-0136}}


\subtitle{III. A relation between $\gamma$-velocities and
$\gamma$-asymmetries}
\titlerunning{High resolution spectroscopy for Cepheids distance
determination III.}
\authorrunning{N. Nardetto and collaborators}

\author{N. Nardetto \inst{1}, A. Stoekl \inst{2}, D. Bersier \inst{3}, T. G. Barnes\inst{4,5} }

\institute{Max-Planck-Institut f\"ur Radioastronomie, Auf dem H\"ugel 
69, 53121 Bonn, Germany \and CRAL, Universit\'e de Lyon, CNRS
(UMR5574), \'Ecole Normale Sup\'erieure de Lyon, F-69007 Lyon,
France  \and Astrophysics Research Institute, Liverpool John Moores
University, Twelve Quays House, Egerton Wharf, Birkenhead, CH41 1LD,
UK \and University of Texas at Austin, McDonald Observatory, 1
University Station, C1402, Austin, TX 78712-0259, USA \and currently
on assignment to the National Science Foundation, 4201 Wilson
Boulevard, Arlington, VA 22230, USA}

\date{Received ... ; accepted ...}

\abstract{Galactic Cepheids in the vicinity of the Sun have a
residual line-of-sight velocity, or $\gamma$-velocity, which shows a
systematic blueshift of about 2~\kms compared to an axisymmetric
rotation model of the Milky Way. This term is either related to the
space motion of the star and, consequently, to the kinematic
structure of our Galaxy, or it is the result of the dynamical
structure of the Cepheids' atmosphere. } {We aim to show that these
residual $\gamma$-velocities are an intrinsic property of Cepheids.}
{We observed nine galactic Cepheids with the HARPS\thanks{High
Accuracy Radial velocity Planetary Search project developed by the
European Southern Observatory} spectroscope, focusing specifically
on 17 spectral lines. For each spectral line of each star, we
computed the $\gamma$-velocity (resp. $\gamma$-asymmetry) as an
average value of the interpolated radial velocity (resp. line
asymmetry) curve. } {For each Cepheid in our sample, a linear
relation is found between the $\gamma$-velocities of the various
spectral lines and their corresponding $\gamma$-asymmetries, showing
that residual $\gamma$-velocities stem from the intrinsic properties
of Cepheids. We also provide a physical reference to the stellar
$\gamma$-velocity: it should be zero when the $\gamma$-asymmetry is
zero. Following this definition, we provide very precise and
physically calibrated estimates of the $\gamma$-velocities for all
stars of our sample [in \kms]: $-11.3 \pm 0.3$ [R~TrA], $-3.5 \pm
0.4$ [S~Cru], $-1.5 \pm 0.2$ [Y~Sgr], $9.8 \pm 0.1$ [$\beta$~Dor],
$7.1 \pm 0.1$ [$\zeta$~Gem], $24.6 \pm 0.4$ [RZ~Vel], $4.4 \pm 0.1$
[$\ell$~Car], $25.7 \pm 0.2$ [RS~Pup]. Finally, we investigated
several physical explanations for these $\gamma$-asymmetries like
velocity gradients or the relative motion of the line-forming region
compared to the corresponding mass elements. However, none of these
hypotheses seems to be entirely satisfactory to explain the
observations. } {To understand this very subtle $\gamma$-asymmetry
effect, further numerical studies are needed. Cepheids' atmosphere
are strongly affected by pulsational dynamics, convective flows,
nonlinear physics, and complex radiative transport. Hence, all of
these effects have to be incorporated simultaneously and
consistently into the numerical models to reproduce the observed
line profiles in detail. }


\maketitle

\section{Introduction}\label{s_Introduction}

Cepheids are very important astrophysical objects due to their
well-known period-luminosity (\emph{PL}) relation. Based on this
relation, a multi-decade work allowed us to determine the kinematic
structure of the Milky Way (in particular its rotation) and to reach
cosmologically significant extragalactic distances (see Hubble Space
Telescope Key Project, Freedman et al.\ 2001). In the second paper
of this series, Nardetto et al.\ (2007) (hereafter Paper~II)
established a clear link between the distance scale and the
dynamical structure of Cepheids' atmosphere through a
period-projection factor (\emph{Pp}) relation. Similarly, studies
concerning the kinematics of the Milky Way might be closely related
to the dynamical structure of Cepheids' atmosphere.

Concerning the distance scale, near-infrared interferometry
currently provides a new, quasi-geometrical way to determine the
distance of galactic Cepheids up to $1$~kpc (see e.g. Sasselov \&
Karovska \cite{Sasselov94} and Kervella et al.\ \cite{Kervella04}).
The basic principle of the Interferometric Baade-Wesselink method
(IBW) is to compare the linear and angular size {\it variation} of a
pulsating star in order to derive its distance through a simple
division. The key point is that interferometric measurements {\it in
the continuum} lead to angular diameters corresponding to the {\it
photospheric} layer, while the linear stellar radius variation is
deduced by spectroscopy, i.e., based on line-forming regions which
form higher in the atmosphere. Thus, radial velocities
$V_{\mathrm{rad}}$, which are derived from line profiles, include
the integration in two directions: over the stellar surface through
limb-darkening and over the atmospheric layers through velocity
gradients. All these phenomena are currently merged in one specific
quantity, generally considered constant with time: the projection
factor $p$, defined as $V_{\mathrm{puls}}=pV_{\mathrm{rad}}$, where
$V_{\mathrm{puls}}$ is defined as the {\it photospheric} pulsation
velocity (Nardetto et al. 2004). $V_{\mathrm{puls}}$ is then
integrated with time to derive the photospheric radius variation.
The precision in the distance currently obtained with the IBW method
is a few percent. However, it remains strongly dependent on the
projection factor. If a constant projection factor is used
(generally $p=1.36$ for all stars) to derive the \emph{PL} relation,
errors of $0.10$ and $0.03$ on the slope and zero-point of the
\emph{PL} relation can be introduced. This means that distances can
be overestimated by 10\% for long-period Cepheids (Paper~II).

In Paper~II, we divide the projection factor into three
sub-concepts: (1) a geometrical effect related to the
limb-darkening, (2) the velocity gradient within the atmosphere, and
(3) the differential motion of the ``optical'' pulsating photosphere
compared to the corresponding mass elements, called
$f_{\mathrm{o-g}}$. Even if the \emph{Pp} relation was recently
confirmed by HST observations (Fouqu\'e et al.\ 2007), the
$f_{\mathrm{o-g}}$ is relatively uncertain and currently entirely
based on hydrodynamical calculations. Nevertheless, a key point is
that this quantity should be, in principle, related to the so-called
$\gamma$-velocity term. Indeed, as already mentioned by Sabbey et
al.\ (1995):

``The observed unequal line asymmetry magnitudes during
contraction and expansion [also observed in the first paper of this
series, Nardetto et al.\ 2006, hereafter Paper~I], are due to the
varying depth of spectral line formation over a Cepheid pulsation
cycle. In other words, the photospheric spectral lines in a Cepheid
atmosphere are not associated with the same gas particles during
cycle. Therefore, they are not required to, and indeed do not,
comply with {\it path conservation}

$$ \int V_{\mathrm{rad}}d\phi=0$$

where the integral is over a complete Cepheid cycle. However path
conservation is a basic assumption in the BW method (Gautschy
1987).''

The state of the art of the BW methods, concerning the
$\gamma$-velocity is the following: The $\gamma$-velocity is
generally removed, i.e.\ the average value of the radial velocity
curve is forced to zero before integrating. However, the
differential motion between the line-forming region and the gas also
modifies the velocity amplitude. The $f_{\mathrm{o-g}}$ was
introduced, based on hydrodynamical models in Paper~II, in order to
correct this velocity amplitude effect. The next step is to find a
correlation between $f_{\mathrm{o-g}}$ and the $\gamma$-velocity. In
this paper, we will provide some indications, but the problem
remains unsolved.

The $\gamma$-velocities are also of great importance for the
determination of the kinematic structure of the Milky Way. Galactic
Cepheids in the solar vicinity show a residual line-of-sight
velocity (a ``K-term'') in their radial velocities which is
systematically blueshifted of about 2~\kms\,compared to an
axisymmetric rotation model of the Milky Way (Camm 1938, 1944;
Parenago 1945; Stibbs 1956; Wielen 1974; Caldwell \& Coulson 1987;
Moffett \& Barnes 1987; Wilson et al.\ 1991; Pont, Mayor \& Burki
1994). Wielen (1974) found no correlation of the K-term with any
obvious parameter such as period, amplitude or distance, and
concluded that the K-term is an intrinsic property of Cepheid
atmospheres. On the other hand, Pont et al.\ (1994) tried to revive
Camm (1944) and Parenago's (1945) suggestion that the K-term is due
to a real effect in the dynamics of the Galaxy. In addition, Butler
et al.\ (1996) found a $\gamma$-velocity reduced by 2~\kms\,by
introducing velocity gradients in hydrostatic stellar atmospheres
models. The K-term problem is still a matter of debate today.

Based on very high quality HARPS observations and careful
methodology (Sect.~\ref{s_observation}), we will show that
$\gamma$-velocities are due to intrinsic properties of Cepheids
(Sect.~\ref{s_relation}). Finally, we discuss our results in
Sect.~\ref{s_discussion}.

\begin{figure}[htbp]
\begin{center}
\resizebox{0.8\hsize}{!}{\includegraphics[clip=true]{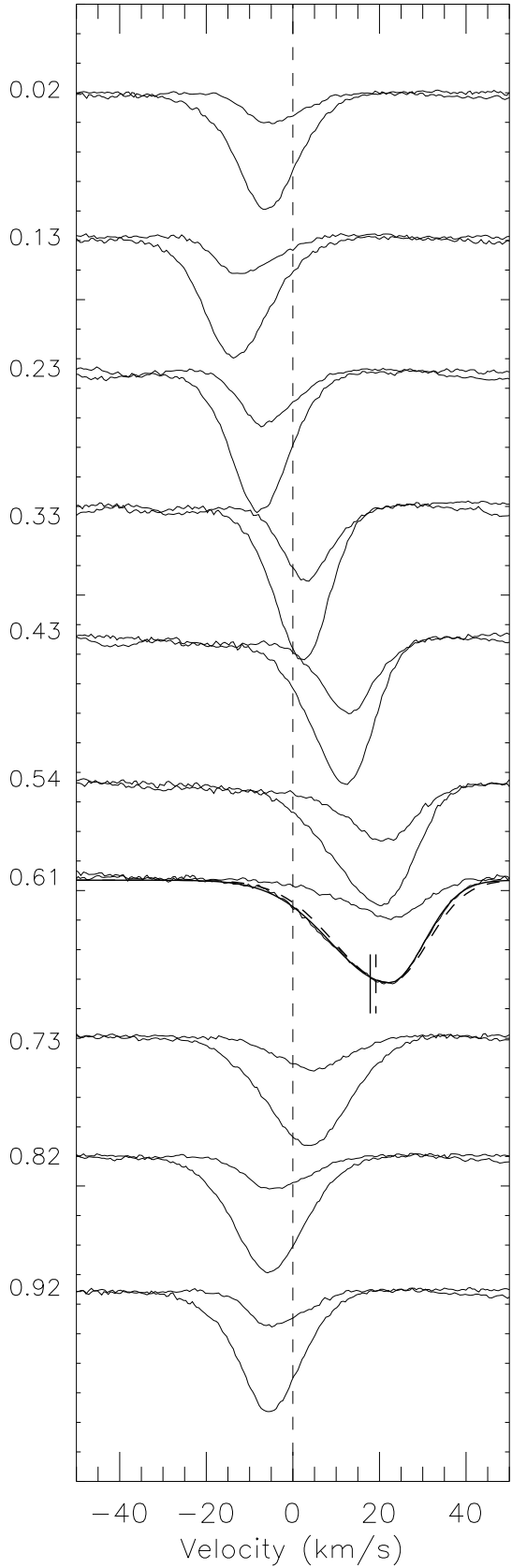}}
\caption{\ion{Fe}{I} 4896.4 ($D \simeq 8$\%) and \ion{Fe}{I} 6024.1
($D \simeq 30$\%) spectral lines evolution of $\beta$ Dor. Pulsation
phases are given on the left of each profile. Wavelengths have been
translated into velocities for comparison (positive velocities
correspond to a redshift, motion toward us). The systematic
difference in lines asymmetries and velocities is clear. The solid
and dashed lines at $\phi=0.61$ show that modifying artificially the
line asymmetry induces a change in the resulting centroid velocity
(see Sect. \ref{s_discussion}).} \label{Fig_film}
\end{center}
\end{figure}

\section{HARPS observations} \label{s_observation}

Ten stars have been observed with the HARPS spectrometer
($R=120000$): \object{R~Tra}, \object{S~Cru}, \object{Y~Sgr},
\object{$\beta$~Dor}, \object{$\zeta$~Gem}, Y~Oph, \object{RZ~Vel},
\object{$\ell$~Car}, \object{RS~Pup} and X~Sgr. X~Sgr is an atypical
Cepheid presenting several components in the spectral line profiles.
It was studied separately by Mathias et al.\ (2006). Y~Oph is not
studied here in detail due to its insufficient phase coverage (see
Paper~II, Fig. 3). We thus consider 8 Cepheids in this paper.

Using Kurucz models \cite{kurucz92} we have identified about 150
unblended spectral lines. In Paper~II, we carefully selected 17
spectral lines following two criteria: (1) the continuum must to be
perfectly defined for all pulsation phases of all stars, in order to
avoid bias in the determination of the line depth, (2) the selected
sample of lines has to cover a large range of depth. The spectral
lines selected are presented in Tab.~1 of Paper~II.

As in Paper~I, we use bi-Gaussian fits to derive line asymmetries.
We repeat here the main equations in order to show that there is no
{\it a priori} link between the line asymmetry and our
 {\it centroid} method of the radial velocity determination.
The centroid radial velocity ($RV_{\mathrm c}$), or the first moment
of the spectral line profile, has been estimated as
\begin{equation} \label{Eq_CDG}
 RV_{\mathrm c} = \frac{\int_{\rm line} \lambda S(\lambda) d\lambda}{\int_{\rm line} S(\lambda) d\lambda}
\end{equation}
where $S(\lambda)$ is the {\it observed} line profile. Then, the
radial velocity corresponding to the minimum pixel, full width at
half-maximum (FWHM), and asymmetry are derived simultaneously by
applying a classical $\chi^{2}$ minimization algorithm between the
observed line profile ($S(\lambda)$) and a modeled spectral line
profile ($f(\lambda)$). The corresponding reduced $\chi^2$ is
\begin{equation} \label{Eq_X2_BiGaussian}
{\chi_{\mathrm red}}^{2} = \frac{1}{N-\nu}
\sum_{i=0}^{N}{\frac{(S(\lambda_i)-f(\lambda_i))^{2}}{\sigma(\lambda_i)^2}}
\end{equation}
with $N$ being the number of pixels in the spectral line, $\nu$ the
number of degrees of freedom and $\sigma(\lambda_i)=$ {\it SNR}
$*f(\lambda_i)$ the statistical uncertainty associated to each
pixel. {\it SNR} is the estimate of the signal-to-noise Ratio in the
continuum.

The analytic line profile is defined by
\begin{equation} \label{Eq_BiGaussian1}
f(\lambda)= 1 - D \exp \left( \frac{4 \ln 2 (\lambda -
\lambda_{\mathrm m}) ^2}{(FWHM(1+A))^2} \right) \mbox{ if } \lambda
> \lambda_{\mathrm{m}}
\end{equation}
and
\begin{equation} \label{Eq_BiGaussian2}
 f(\lambda)= 1  - D \exp \left( \frac{4 \ln 2 (\lambda - \lambda_{\mathrm m})
^2}{(FWHM(1-A))^2} \right) \mbox{ if } \lambda <
\lambda_{\mathrm{m}}
\end{equation}
with four free parameters: $D$, the depth of the line
(dimensionless); $\lambda_{\mathrm m}$, the wavelength associated to
the minimum of the line (in \AA); {\it FWHM}, the Full-Width at
Half-Maximum in the line (in \AA); and $A$, the line asymmetry
relative to the $FWHM$ (in \%). Online tables (Nardetto et al.\
2007) present the resulting values of $RV_{\mathrm{m}}$,
$RV_{\mathrm{c}}$, $FWHM$, $D$, $A$, $SNR$, and $\chi_{red}^{2}$
together with the corresponding uncertainties computed from the
fitting method. In the following, we will only use $RV_{\mathrm c}$
and $A$, which are clearly \emph{independent} given the manner of
their determinations.

We insist on the $RV_{\mathrm{c}}$ definition of the radial velocity
since it is absolutely required to allow direct comparisons between
$\gamma$-velocities of different spectral lines from different
Cepheids. Indeed, it is the {\it only} method which provides a
radial velocity independent of the rotation (projected on the line
of sight) and the natural width of the spectral line (Burki et al.\
1982 and Paper~I).

The $RV_{\mathrm{c}}$ and $A$ quantities for 17 selected spectral
lines for each of the 8 stars have been interpolated over the
pulsation phase using a periodic cubic spline function. The
interpolation is performed either directly on the observational
points (e.g.\ $\beta$~Dor) or using arbitrary pivot points (e.g.\
RZ~Vel). In the latter case, a classical minimization process
between the observations and the interpolated curve is used to
optimize the position of the pivot points. All interpolated curves
presented in this study were derived using one of these two methods.
Finally, $A_{\gamma}$ and $V_{\gamma}$ are calculated by
\emph{averaging} the $A(\phi)$ and $RV_{\mathrm{c}}(\phi)$
interpolated curves, where $\phi$ is the pulsation phase. The
corresponding uncertainties are defined as the average values of
individual uncertainties on the observational points.

\begin{figure*}[htbp]
\resizebox{\hsize}{!}{\includegraphics[clip=true]{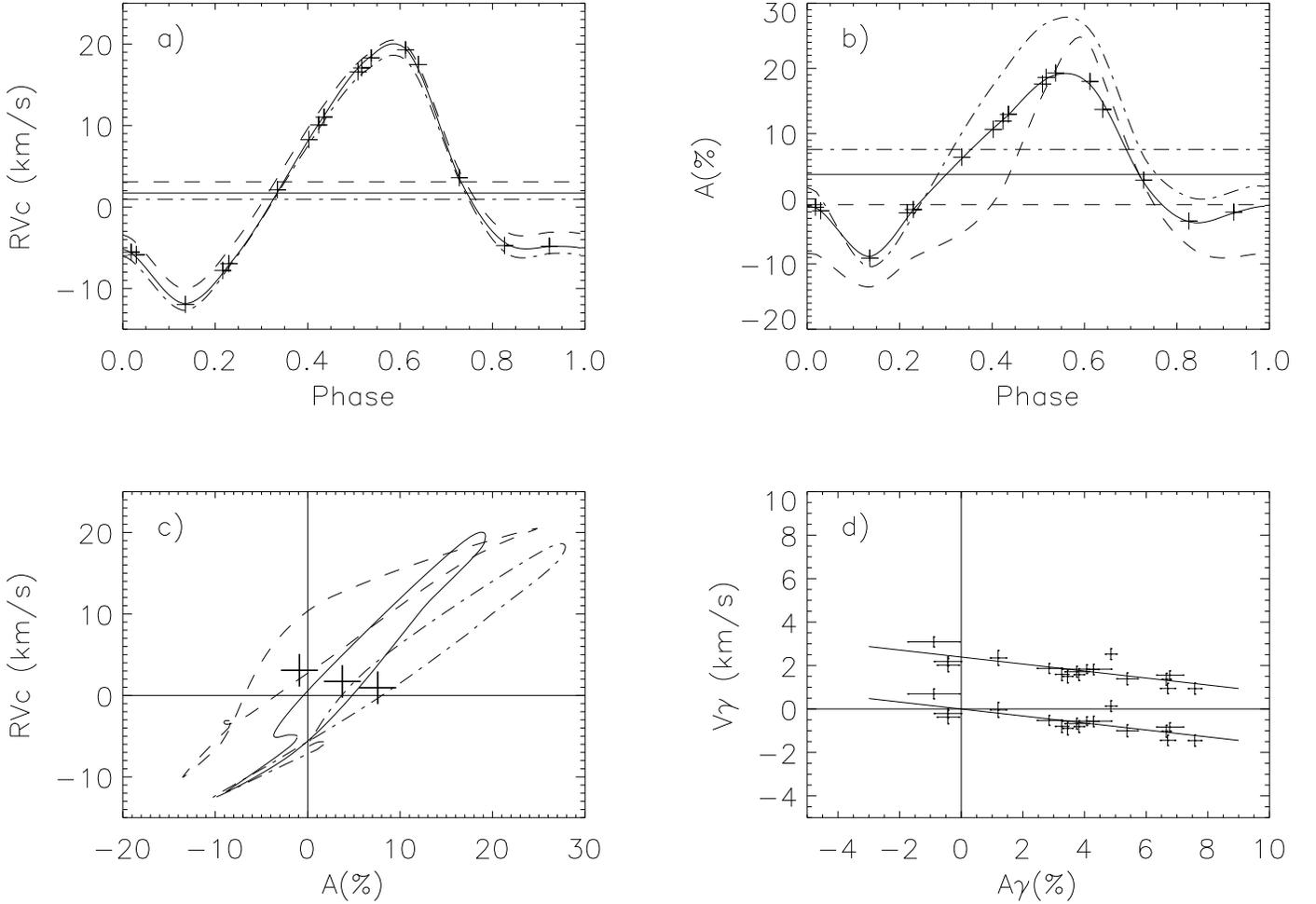}}
\caption{ $RV_{\mathrm{c}}$ (a) and $A$ (b) are represented as a
function of the pulsation phase for three spectral lines in the case
of $\beta$~Dor: \ion{Fe}{I} 4896.4 (dashed line), \ion{Fe}{I} 5373.7
(solid line), and \ion{Fe}{I} 6024.1 (dot-dashed line). The
$RV_{\mathrm{c}}$ curves include only the GCD $\gamma$-velocity
correction. The actual measurements (crosses) are only indicated in
the case of the \ion{Fe}{I} 5373.7 line. The corresponding
uncertainties are too small to be visible in the plot. Horizontal
lines correspond to the average values $V_{\gamma}$ and $A_{\gamma}$
of the interpolated curves, respectively. (c) $RV_{\mathrm{c}}-A$
plots and the corresponding ($A_{\gamma}$, $V_{\gamma}$) average
values (crosses) for the three different lines. Although the
$RV_{\mathrm{c}}-A$ plots have different shapes, the average values
are aligned. (d) Generalization of diagram (c) for all spectral
lines. The $RV_{\mathrm{c}}-A$ plots are not included for clarity.
The upper values are without any correction except the GCD
$\gamma$-velocity. The origin of the plot is then used as a physical
reference for all spectral line $\gamma$-velocities of the star
(lower values). The $V_{\gamma} A_{\gamma}$ correlation found is a
strong indication that $\gamma$-velocities are related to intrinsic
physical properties of Cepheids' atmospheres.} \label{Fig_betaDor}
\end{figure*}

\begin{figure*}[htbp]
\resizebox{\hsize}{!}{\includegraphics[clip=true]{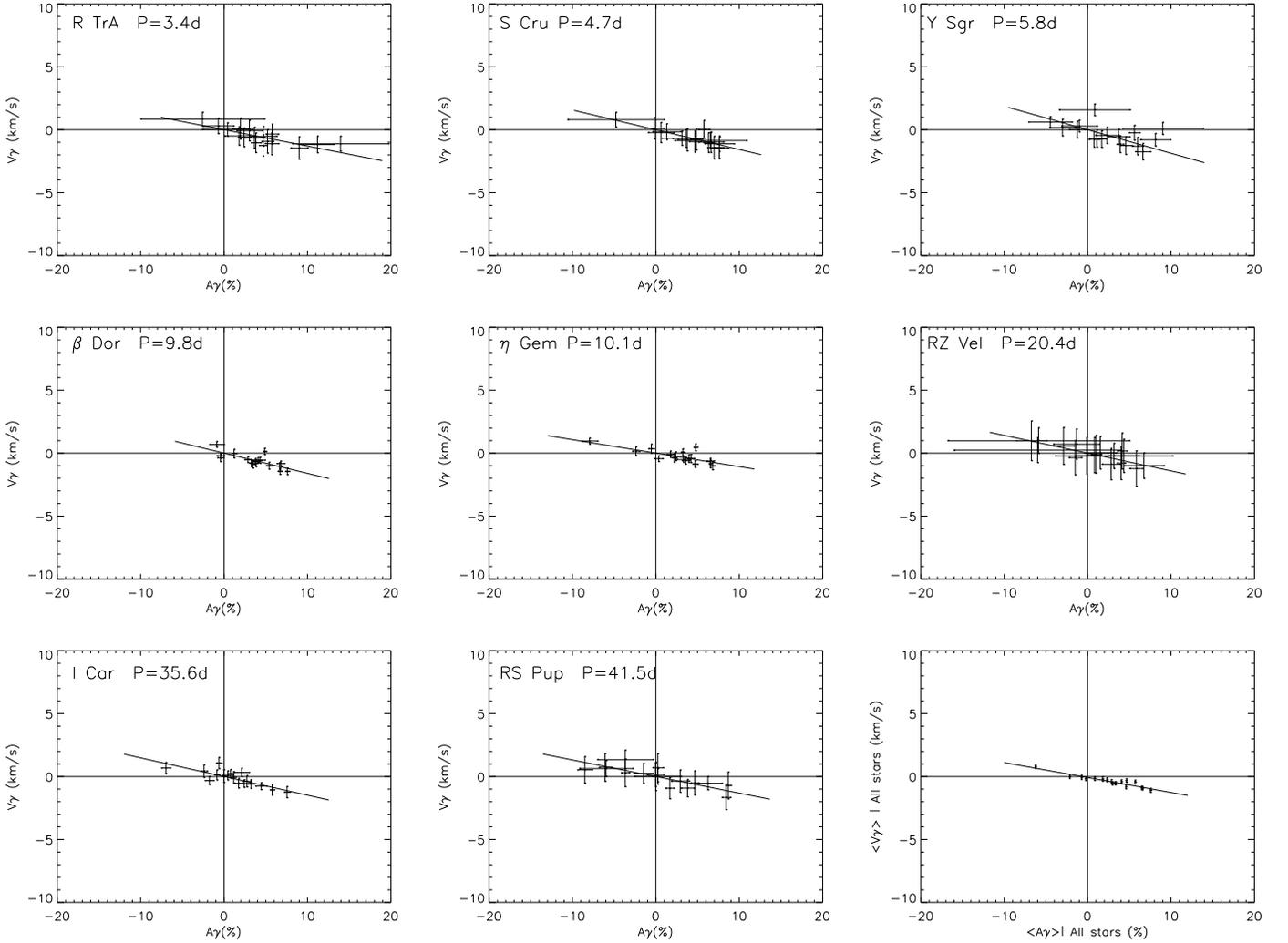}}
\caption{Same as Fig.~\ref{Fig_betaDor}d for all Cepheids of our
sample. The origin of the $V_{\gamma}A_{\gamma}$ plots are used as
physical references. The correction applied ($V_{\gamma
\star}=V_{\gamma}\mathrm{GCD}+b_{\mathrm{0}}$), together with the
slope values ($a_{\mathrm{0}}$), are indicated in
Tab.~\ref{Tab_results}. In the last panel an weighted average is done
over all stars. The resulting $V_{\gamma}A_{\gamma}$ linear curve is
very precise ($a_{\mathrm{0}}=0.15 \pm 0.01$).} \label{Fig_gamma}
\end{figure*}

\begin{table*}
\begin{center}
\caption[]{Linear relations between the $\gamma$-velocities and the
$\gamma$-asymmetries $ V_{\gamma} = a_{\mathrm{0}}  A_{\gamma} +
b_{\mathrm{0}}$ for all stars. The subscripts give the $1\sigma$
uncertainty. The {\it reduced} $\chi^2$, defined as
$\chi_{\mathrm{red}}^2=\frac{\chi^2}{N-\nu}$ with N being the number
of spectral lines and $\nu$ the number of degrees of freedom, is
also indicated. The radial velocities are initially corrected from
the GCD $\gamma$-velocities, and then a second correction
$b_{\mathrm{0}}$ is applied, taking the origin of
$A_{\gamma}V_{\gamma}$ plots as a reference: $V_{\gamma}=0$ implies
$A_{\gamma}=0$. The last column corresponds to the final physically
calibrated $\gamma$-velocities for each star.}\label{Tab_results}
\begin{tabular}{lccccccc}
\hline \hline \noalign{\smallskip}

Name     &  HD     &  $P^{\mathrm{\tiny (a)}}$   &   $a_{\mathrm{0}}$ &       $\chi_{\mathrm{red}}^2$        & $V_{\gamma \mathrm{GCD}}$ & correction $b_{\mathrm{0}}$ & $V_{\gamma \star}$ \\
         &         &  [days]            &           [\kms /\%]                      &                                      &      [\kms]                   &                              [\kms] &      [\kms]             \\
\hline
R TrA       & 135592  &   $3.38925$     &   $   -0.13   _{\pm   0.05    }$  &   3   &   $-13.2$   &   $   1.9 _{\pm   0.3 }$  &   $   -11.3   _{\pm   0.3 }$  \\
S Cru       & 112044  &   $4.68976$     &   $   -0.16   _{\pm   0.07    }$  &   2   &   $-7.1$    &   $   3.6 _{\pm   0.4 }$  &   $   -3.5   _{\pm   0.4 }$  \\
Y Sgr       & 168608  &   $5.77338$     &   $   -0.19   _{\pm   0.04    }$  &   14  &   $-2.5$    &   $   1.0 _{\pm   0.2 }$  &   $   -1.5    _{\pm   0.2 }$  \\
$\beta$ Dor & 37350   &   $9.84262$     &   $   -0.16   _{\pm   0.02    }$  &   30  &   $7.4$     &   $   2.4 _{\pm   0.1 }$  &   $   9.8 _{\pm   0.1 }$  \\
$\zeta$ Gem & 52973   &   $10.14960$    &   $   -0.11   _{\pm   0.02    }$  &   22  &   $6.9$     &   $   0.2 _{\pm   0.1 }$  &   $   7.1 _{\pm   0.1 }$  \\
RZ Vel      & 73502   &   $20.40020$    &   $   -0.14   _{\pm   0.09    }$  &   1   &   $24.1$    &   $   0.5 _{\pm   0.4 }$  &   $   24.6    _{\pm   0.4 }$  \\
$\ell$~Car  & 84810   &   $35.55134$    &   $   -0.15   _{\pm   0.03    }$  &   15  &   $3.6$     &   $   0.8 _{\pm   0.1 }$  &   $   4.4 _{\pm   0.1 }$  \\
RS Pup      & 68860   &   $41.51500$    &   $   -0.13   _{\pm   0.05    }$  &   4   &   $22.1$    &   $   3.6 _{\pm   0.2 }$  &   $   25.7    _{\pm   0.2 }$  \\

\hline \noalign{\smallskip}
\end{tabular}
\end{center}
\begin{list}{}{}

\item[$^{\mathrm{a}}$] The corresponding Julian dates
($T_{\mathrm{o}}$) can be found in Paper~II.

\end{list}
\end{table*}

\section{A relation between $\gamma$-velocities and $\gamma$-asymmetries} \label{s_relation}

The aim of this section is to study the $\gamma$-velocities and to
show that they consist of two components: one related to the space
motion of the star itself, and one (the K-term) related to the
dynamical structure of Cepheids' atmosphere.

\subsection{Methodology presented in the case of $\beta$~Dor}

To introduce our methodology, we discus the exemplary case of
$\beta$~Dor in detail.

First, interpolated radial velocity curves derived from all spectral
lines were corrected by the $\gamma$-velocity found in the Galactic
Cepheid Database of the David Dunlap
Observatory\footnote{http://www.astro.utoronto.ca/DDO/research/cepheids/}
(Fernie et al.\ 1995, hereafter $V_{\gamma \mathrm{GCD}}$, see
Tab.~\ref{Tab_results}).

Fig.~\ref{Fig_film} shows the spectral line profile of two metallic
lines \ion{Fe}{I} 4896.4 ($D \simeq 8$\% of the continuum) and
\ion{Fe}{I} 6024.1 ($D \simeq 30$\%) as a function of the pulsation
phase. We translated wavelengths into velocities for comparison. We
used spectral lines with different line depths for clarity, but the
results are actually {\it independent} of the line depth. Two
qualitative key-points concerning Fig.~\ref{Fig_film} should be
mentioned: (1) the \ion{Fe}{I} 4896.4 spectral line (smaller line
depth) seems to be systematically redshifted compared to the
\ion{Fe}{I} 6024.1 spectral line (look, for instance, at the pixel
minimum of each profile) and (2) its asymmetry is also
systematically larger (in absolute value) from phase $\phi=0.82$ to
$\phi=0.23$, while it is systematically lower (in absolute value)
from phase $\phi=0.33$ to $\phi=0.61$. This leads us to the idea of
a correlation between the $\gamma$-velocity and the
$\gamma$-asymmetry. In Fig.~\ref{Fig_betaDor}ab we present the
corresponding interpolated radial velocity and asymmetry curves for
these two spectral lines. Another line of intermediate depth is also
presented (\ion{Fe}{I} 5373, solid line). From the interpolated
curves, we now calculate the $\gamma$-velocities and -asymmetries
corresponding to each spectral line (horizontal lines). An
anti-correlation is clearly seen.

In Fig.~\ref{Fig_betaDor}c, $RV_{\mathrm{c}}-A$ plots and the
corresponding ($A_{\gamma}$, $V_{\gamma}$) average values (crosses)
for the three different lines are presented. Although the
$RV_{\mathrm{c}}-A$ plots have different shapes (and this is
confirmed for all spectral lines), which is the result of the
dynamical structure of the Cepheid atmosphere, the average values
(big crosses) are aligned, confirming the correlation found in
Fig.~\ref{Fig_betaDor}ab.

Fig.~\ref{Fig_betaDor}d is a generalization of diagram (c) for all
spectral lines. The $RV_{\mathrm{c}}-A$ curves are not included for
clarity. Upper values correspond to residual $\gamma$-velocities
$V_{\gamma}(i)$ of the 17 selected spectral lines $i$ after the GCD
$\gamma$-velocity correction. A linear fit is performed, and we find
the relation
$$ V_{\gamma}(i)-V_{\gamma \mathrm{GCD}}=
a_{\mathrm{0}} A_{\gamma}(i) + b_{\mathrm{0}},$$ with
$a_{\mathrm{0}}=-0.16 \pm 0.02$ [\kms per \%] and
$b_{\mathrm{0}}=2.4 \pm 0.1$ [\kms]. The origin of the plot is taken
as a reference: the $\gamma$-velocity is assumed to be zero when the
$\gamma$-asymmetry is zero. This means that all points
$$ (A_{\gamma}(i), V_{\gamma}(i)-V_{\gamma \mathrm{GCD}} )$$
are translated into
$$ (A_{\gamma}(i), V_{\gamma}(i)-(V_{\gamma \mathrm{GCD}}+b_{\mathrm{0}}) )$$
which allows the definition of a physically calibrated
$\gamma$-velocity for $\beta$ Dor
$$ V_{\gamma\star}=V_{\gamma \mathrm{GCD}}+b_{\mathrm{0}}=7.4+2.4=9.8_{\pm 0.1} \:\mathrm{km\: s}^{-1} \, .$$
This quantity is very precise due to the very high S/N and spectral
resolution of HARPS data.

In principle, the line asymmetry and the $\gamma$-asymmetry are
supposed to be the result of the dynamical structure of the Cepheid
atmosphere {\it only}. The $V_{\gamma}(i) A_{\gamma}(i)$ (hereafter
$V_{\gamma}A_{\gamma}$) correlation found is then a strong
indication that residual $\gamma$-velocities (after correction by
the GCD $\gamma$-velocities) are related to intrinsic physical
properties of Cepheids' atmosphere and not to a real effect in
dynamics of the Galaxy (see discussion).

\subsection{$V_{\gamma}A_{\gamma}$ linear curves of all stars of our sample}

$V_{\gamma}$ and $A_{\gamma}$ were derived for all spectral lines
and for all stars of our sample using the same method as presented
in the case of $\beta$~Dor.

Tab.~\ref{Tab_results} gives $V_{\gamma \mathrm{GCD}}$, the slope
$a_{\mathrm{0}}$ of the interpolated $V_{\gamma} A_{\gamma}$ curves,
the correction $b_{\mathrm{0}}$ applied, and our final
$\gamma$-velocities $V_{\gamma \star}$ for each star. The
$V_{\gamma} A_{\gamma}$ plots are shown in Fig.~\ref{Fig_gamma}.
Linear correlation curves between $V_{\gamma}$ and $A_{\gamma}$ are
found for all stars of our sample, and our $b_{\mathrm{0}}$
corrections range from 0.2 to 3.6 \kms. The average value is $1.8
\pm 0.2$ \kms, which is consistent with the 2~\kms ``K-term'' found
in the literature. We discuss these important observational results
in the next section.


\section{Discussion}\label{s_discussion}

\subsection{$\gamma$-velocities}

The linear relation between $A_{\gamma}$ and $V_{\gamma}$ can be
easily understood. The basic principle is demonstrated in
Fig.~\ref{Fig_film} for the pulsation phase $\phi=0.61$. For
clarity, we only present the argument for one pulsation phase.

The solid line is the bi-Gaussian fit of the \ion{Fe}{I} 6024.1
spectral line. Artificially decreasing the bi-Gaussian asymmetry of
this spectral line by the \emph{average} asymmetry of the
\ion{Fe}{I} 6024.1 line, $A_{\gamma} = 7.58\%$, as shown in
Fig.~\ref{Fig_film} with a dashed-line, we find that the
centroid-velocity increases (i.e.\ is redshifted) by an amount of
+1.45~\kms.  To illustrate this, we included in Fig.~\ref{Fig_film}
the computed line positions $RV_{\mathrm{c}}$ by vertical lines
before (solid line) and after (dot-dashed line) modification of the
bi-Gaussian asymmetry. Generally, we find that forcing the average
line asymmetry $A_{\gamma}$ to zero by uniformly changing the line
asymmetry \emph{by the same amount} at all phases alters the derived
centroid velocities $RV_{\mathrm{c}}$ in such a way that the
resulting $\gamma$-velocity becomes zero. For instance, the point
($A_{\gamma}=7.58$\%, $V_{\gamma}=1.45~$\kms) of the
$V_{\gamma}A_{\gamma}$ linear curve of $\beta$~Dor (Fig.
\ref{Fig_gamma}) translates into ($A_{\gamma}=0$\%,
$V_{\gamma}=0~$\kms).

This result is obtained universally, regardless which spectral line
or star is considered. Basically, we can conclude that the
$\gamma$-velocities are a side-effect of the line asymmetries and
related to the problem of determining line positions of asymmetric
lines. This view is also supported by the high conformity of the
slopes $a_{\mathrm{0}}$ of the $V_{\gamma}A_{\gamma}$ relation for
all sample stars (see Tab.~\ref{Tab_results}) which (almost) agree
within $1\sigma$ uncertainties.

Although this method of arbitrarily modifying the line asymmetry is
a rather \emph{ad hoc} method, it still allowed us to gain some
insight into the origin of the $\gamma$-velocities. In particular,
it shows (1) that the residual $\gamma$-velocities are related to
the shapes of the spectral lines and, consequently, to an intrinsic
property of Cepheids and (2) that the main physical question is not
to understand the $\gamma$-velocities, but to understand the
$\gamma$-asymmetries.

Another important point is that this interpretation provides a
physical meaning to the $\gamma$-velocities and, hence, a physical
\emph{reference}. The relation between $\gamma$-asymmetry and
$\gamma$-velocity allows us to compute the contribution of the
dynamical Cepheid's atmosphere to the $\gamma$-velocity. We can thus
really provide a $\gamma$-velocity corresponding to the space motion
of the star itself ($V_{\gamma \star}$), independently of the
dynamical structure of its atmosphere. From Tab.~\ref{Tab_results},
we find an \emph{average} systematic red-shift correction of
$b_{\mathrm{0}}=1.8 \pm 0.2$ \kms (averaged over 8 stars) between
our physically calibrated $\gamma$-velocities and the ones found in
the GCD, commonly used by the community. Consequently, the K-term
(blueshifted) found in the literature is not due to the kinematic
structure of the Galaxy, but to a bias in the previous methods of
deriving the $\gamma$-velocities (cross-correlation, Gaussian fit of
the spectral line), most likely due to these $A_{\gamma}V_{\gamma}$
linear relations. By using only one metallic line to derive the
$\gamma$-velocity, one can make, for instance, an error (or find
inconsistencies) ranging from -2 to +1 \kms. Finally, our results
are thus consistent with an axisymmetric rotation model of the Milky
Way.

\subsection{$\gamma$-asymmetries}

Different aspects of the $\gamma$-asymmetries must be pointed out.

\begin{figure}[htbp]
\begin{center}
\resizebox{0.95\hsize}{!}{\includegraphics[clip=true]{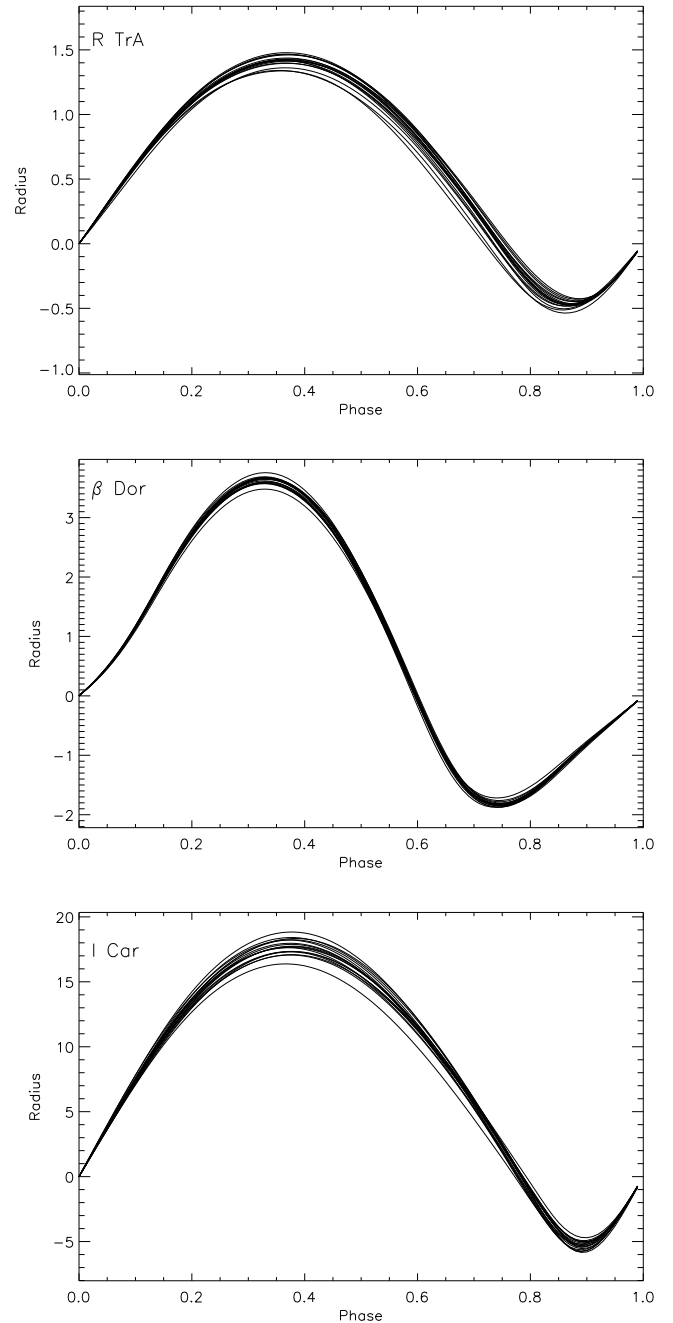}}
\caption{Radius (in solar radii) as a function of the pulsation
phase for spectral lines of our sample. The radial velocity curves
have been corrected from their $\gamma$-velocities in order to
investigate phase shifts. Important differences are found from one
line to the other, which are mainly due to the velocity gradient
within the atmosphere. Actually, there is a linear relation between
the amplitude of the radius variation and the line depth (see for
instance Fig. 4 of Paper II). This indicates why it is essential to
use \emph{dynamic} projection factors. A phase shift is observed
only for R~TrA and RS~Pup.} \label{Fig_rayon}
\end{center}
\end{figure}

\begin{figure}[htbp]
\begin{center}
\resizebox{0.95\hsize}{!}{\includegraphics[clip=true]{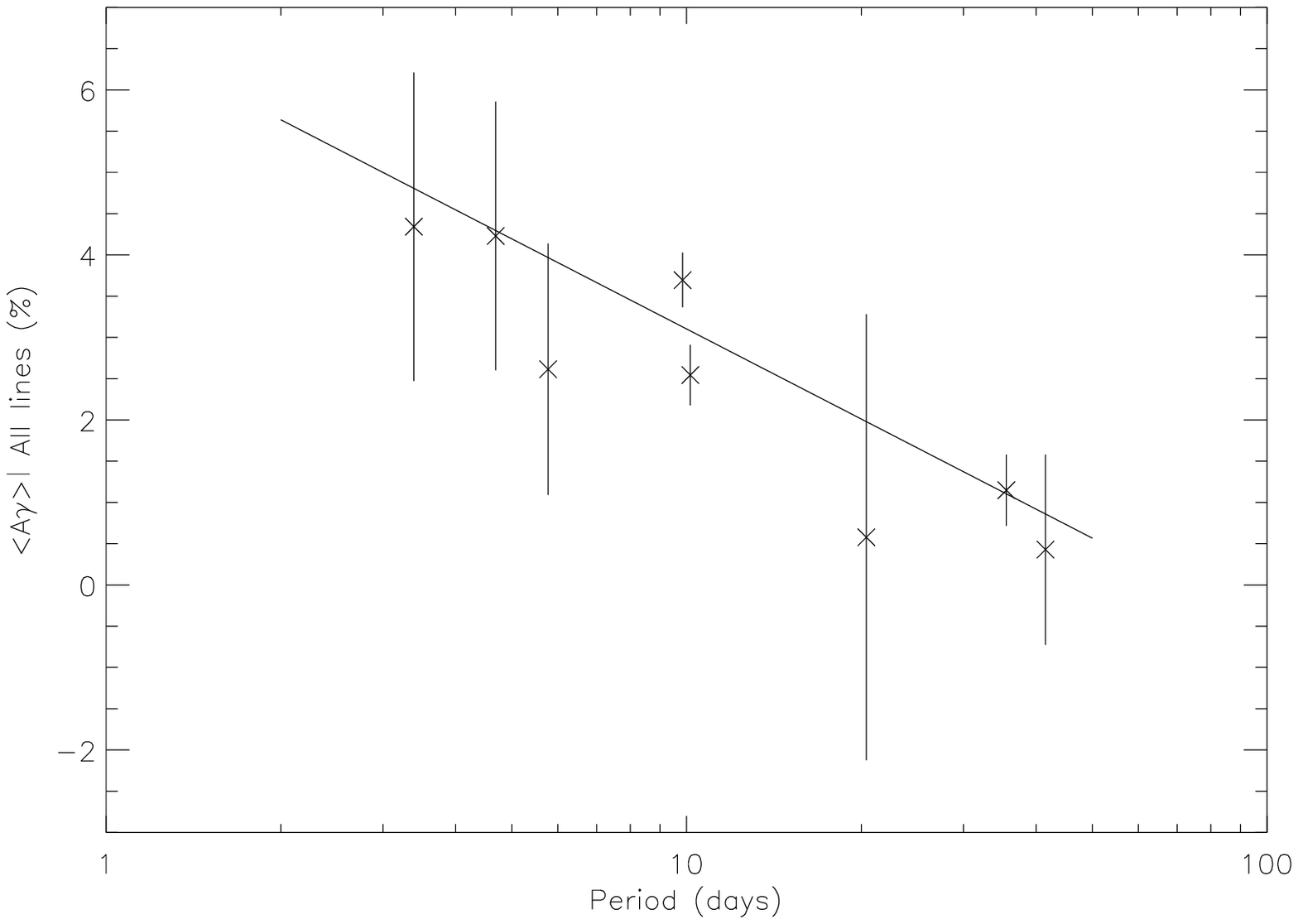}}
\caption{$A_{\gamma}$ averaged (without weighting) over all lines is
given as a function of the pulsation period of the star. The plot is
a generalization of the Fig. 14a of Paper~I.} \label{Fig_asy_p}
\end{center}
\end{figure}

\begin{itemize}

\item The $\gamma$-asymmetries are different from one spectral line to the
other. This was the motivation for our $A_{\gamma}V_{\gamma}$
relations. Moreover, $A_{\gamma}$ can be positive or negative.

\item  Given the precision of our data, we find no particular link between the
line depth (i.e.\ where the line forms within the atmosphere, see
Paper~II) and the $\gamma$-asymmetries. Moreover, no relation was
found between the $\gamma$-asymmetries and the wavelength.

\item In order to investigate phase shifts, which might be related to the $\gamma$-asymmetries, we
integrate radial velocities over time (after correcting by the
corresponding $V_{\gamma}(i)$) and obtain the radius as a function
of the pulsation phase for different spectral lines. Such phase
shifts are observed clearly \emph{only} for R~TrA and RS Pup (see
Fig.~\ref{Fig_rayon}).

\item One can see in Fig.~\ref{Fig_betaDor}b that the difference in asymmetry
between the \ion{Fe}{I} 6024.8\AA\ (dot-dashed line) and \ion{Fe}{I}
4896.4\AA\ (dashed line) asymmetry curves is surprisingly almost
\emph{constant} with the pulsation phase. Even it is not a general
rule (see solid line of Fig.~\ref{Fig_betaDor}b), this effect is
frequent in our data, and seems to be important to understand the
meaning of the $\gamma$-asymmetries.

\item Although the $A(\phi)$ and $RV_{\mathrm{c}}(\phi)$ curves vary
in phase, there is no \emph{strict} correlation between the line
asymmetry $A(\phi)$ and the radial velocity $RV_{\mathrm{c}}(\phi)$
as a function of the pulsation phase $\phi$ (see
Fig.~\ref{Fig_betaDor}c). The slope and shape of the
$RV_{\mathrm{c}}-A$ curves depend on the limb-darkening, the
spectral line width, and the rotation velocity. However, the loop
and the shift in asymmetry ($\gamma$-asymmetry) are mainly due to
the dynamical structure of the Cepheid atmosphere. They cannot be
explained by \emph{static} geometrical models of pulsating stars
(see Paper~I for details).

\item  In the stellar rest frame (defined by $V_{\gamma \star}$), one has
 for a given spectral line and at a specific pulsation phase~$\phi$, $RV_{\mathrm{c}}(\phi)=0$
 while $A(\phi)\ne0$ or, on the
contrary, $A(\phi)=0$ but $RV_{\mathrm{c}}(\phi)\ne0$; i.e.\
$RV_{\mathrm{c}}(\phi)=0$ does not imply $A(\phi)=0$.

\item Interestingly, we find a relation between the $\gamma$-asymmetries and the
pulsation period of Cepheids. In particular, if we define $\langle
A_{\gamma} \rangle$ as the average of the $A_{\gamma}$ quantity over
all spectral lines for a given star, then we find the following
linear relation
$$ \langle A_{\gamma} \rangle
= [-1.6 \pm 0.3]\log P +[6.7 \pm 0.9], $$ where $P$ is the period in
days of the Cepheid. The reduced $\chi^2$ we obtain is 6. This
relation (hereafter $ \langle A_{\gamma} \rangle $) is represented
by Fig.~\ref{Fig_asy_p}. We already presented such a relation in
Paper~I for the \ion{Fe}{I} 6056.005\AA\, spectral line. This new
relation can be considered as a generalization over all spectral
lines of our sample.

\end{itemize}

What could explain the observed $\gamma$-asymmetries? To answer this
question, we investigate three different hypotheses: (1) the
limb-darkening variation with time and within the spectral line, (2)
velocity gradients, and (3) the relative motion of the line-forming
region compared to the corresponding mass elements.

(1) The time- and wavelength dependence of the limb-darkening within
the spectral lines does not explain why the line asymmetry is
\emph{not zero} when the radial velocity is zero. When for instance
the star is at minimum or maximum expansion (i.e.\ radial velocity
is zero), whatever the limb-darkening distribution might be, if
there are no dynamical effects within the atmosphere, then the
spectral line should be symmetric. This is clearly not what we
observe. Consequently, the time- and wavelength dependence of the
limb-darkening may have some effects on the variation of line
asymmetry over the pulsation phase, but it can not be responsible
for all properties mentioned above.

(2) Velocity gradients in the atmosphere can indeed cause an
asymmetric spectral line. Indeed, line asymmetry $A(\phi)$ and
radial velocity $RV_{\mathrm{c}}(\phi)$ basically vary in phase,
even if there is not a strict correlation. This is thus a strong
indication that radial velocity gradients are, to some degree,
responsible for the line asymmetries in these stars. However, they
cannot explain, for instance, the \emph{systematic} shift in
asymmetry (almost constant with the pulsation phase) observed
between the 6024.8\AA\ and \ion{Fe}{I} 4896.4\AA\ spectral line
(Fig.~\ref{Fig_betaDor}). Butler et al.\ (1996) introduced a number
of velocity gradients in their model. They found that ``the effect
of this velocity gradient is to reduce the amplitude of the
pulsational velocity curve at optical depth of $\tau = \frac{2}{3}$
by $20\%$ and to decrease the $\gamma$-velocity by 2~\kms relative
to the standard assumption of a comoving atmosphere''. However, the
main differences they obtain in their velocity curves are near
\emph{extrema}. For instance, in their Fig. 11, all curves (with and
without velocity gradient) vanish at the same pulsation phase. This
is not what we observe in our Fig.~\ref{Fig_betaDor}a, where
\emph{systematic} shifts are obtained. In their study, Butler et
al.\ (1996) applied a ``closure'' constraint: ``It is assumed that
each layer of the stellar atmosphere returns to its starting
position after a pulsation cycle.'' Thus, let us now discuss the
last hypothesis:

(3) The relative motion of the line-forming region with respect to
the corresponding mass elements. When the line-forming region moves
relative to the background atmospherical structure, it will also
move with respect to the background velocity field. Thus, the line
experiences an apparent change in velocity, which has a comparable
effect as a velocity gradient, but with a lower intensity. This
explains why line-forming regions do not comply with the {\it path}
conservation, as discussed in the introduction. However, this cannot
explain, for instance, the \emph{systematic} shift in asymmetry (or
in velocity) observed between the 6024.8\AA\ and \ion{Fe}{I}
4896.4\AA\ spectral lines. Indeed, it would mean that the
line-forming regions corresponding to these two spectral lines have
a \emph{systematic relative} motion ($V_{\gamma}$) compared to the
background, whatever the pulsation cycle considered. Either there
are cycle-to-cycle differences in the {\it path} of the line-forming
regions (which we cannot confirm with our data) or one should invoke
another physical explanation.

\section{ Conclusions }

We found $\gamma$-asymmetries varying from one spectral line to
another, as well as a global dependency on the period of the star
or, correspondingly, on the spatial extension of the Cepheids'
atmosphere. Right now, we have no clear physical explanations for
this effect. Most likely, it results from a combination of several
effects in the dynamical structure of the Cepheids' atmosphere, such
as phase- and wavelength-dependence of the intensity distribution
within the different spectral lines, velocity gradients, non-linear
pulsational effects, shock fronts, and relative motions between
line-forming regions (specific to each spectral line considered) and
the material.


In order to further investigate the line asymmetries, improved
numerical models are required. Since Cepheids' atmospheres are not
in a hydrostatic state but characterized by pulsational dynamics,
one has to perform non-linear, time-dependent simulations of the
underlying pulsation. Snapshots from this temporal evolution
(including the velocity field) can then be used to compute a
detailed frequency-dependent radiative transport. In order to
resolve narrow features such as shock fronts or sharp ionization
regions, high spatial resolution, especially in the line-forming
regions, is needed. Although convective transport plays only a minor
role in the stellar structure of Cepheids in that temperature range
($\simeq 5500 K$, see Tab.~2 in Paper~II), there can still be
considerable convective velocities (some \kms). As a consequence,
the consistent inclusion of the convective velocity field
 -- and of the interaction of convection with pulsation -- in the numerical models
might be crucial for the computation of line asymmetries. Dynamics
in the circumstellar envelope (Kervella et al. 2006; and M\'erand et
al. 2006, 2007) could also be of importance. Confronting such models
with observations (spectral line profiles, spatial- and spectral-
visibility curves from interferometry) may finally lead the way to a
complete picture of the relevant effects in Cepheids' atmospheres.

However, even now, we already have a clear evidence from our
observed linear $A_{\gamma} V_{\gamma}$ relation that the residual
$\gamma$-velocities (or K-term) seen in Cepheids are the result of
the dynamical structure of their atmosphere. This provides a
physical meaning to the $\gamma$-velocities and a physical
reference: \emph{the $\gamma$-velocity should be zero when the
$\gamma$-asymmetry is zero}. This definition of the
$\gamma$-velocities could be used for kinematical studies of the
Galaxy, even though this method requires high signal-to-noise and
high-resolution spectroscopic observations. Using only one metallic
line to derive the $\gamma$-velocity can inflict errors (or
inconsistencies) ranging from -2 to 1 \kms.

\begin{acknowledgements}
Based on observations collected at La Silla observatory, Chile, in
the framework of European Southern Observatory's programs 072.D-0419
and 073.D-0136. We thanks P. Kervella for having provided the HARPS
data and M. Fekety for her careful English correction of the paper.
This research has made use of the SIMBAD and VIZIER databases at
CDS, Strasbourg (France). This material is based in part upon work
by TGB while serving at the National Science Foundation. Any
opinions, findings, and conclusions or recommendations expressed in
this material are those of the authors and do not necessarily
reflect the views of the National Science Foundation. NN
acknowledges the Max Planck Institut for Radioastronomy for
financial support.

\end{acknowledgements}


\begin{table*}
\begin{center}
\caption{$A_{\gamma}$ and $V_{\gamma}$ for all
 spectral lines of all stars. \label{Tab_AgVg}}
\begin{tabular}{lccccccccccc}
\hline \hline \noalign{\smallskip}

Line (\AA)                   &         \multicolumn{2}{c}{R TrA}    &  \multicolumn{2}{c}{S Cru}    &  \multicolumn{2}{c}{Y Sgr}       &   \multicolumn{2}{c}{$\beta$~Dor}    \\
                             &  A$_{\gamma}$[\%] & V$_{\gamma}$[\kms]     &  A$_{\gamma}$[\%]&   V$_{\gamma}$[\kms]       &   A$_{\gamma}$[\%]&   V$_{\gamma}$[\kms]           &     A$_{\gamma}$[\%]&   V$_{\gamma}$[\kms]            \\
    \hline
\ion{Fe}{I}     4683.560    &$      0.40    _{\pm   3.01    }$&$    0.02    _{\pm   0.53    }$&$    3.12    _{\pm   2.41    }$&$    -0.68   _{\pm   0.56    }$&$    -4.49   _{\pm   2.58    }$&$    0.61    _{\pm   0.43    }$&$    -0.44   _{\pm   0.43    }$&$    -0.21   _{\pm   0.21    }$\\
\ion{Fe}{I}     4896.439    &$      -2.58   _{\pm   7.39    }$&$    0.84    _{\pm   0.55    }$&$    -4.80   _{\pm   5.76    }$&$    0.80    _{\pm   0.58    }$&$    0.82    _{\pm   4.23    }$&$    1.58    _{\pm   0.46    }$&$    -0.90   _{\pm   0.84    }$&$    0.70    _{\pm   0.23    }$\\
\ion{Fe}{I}     5054.643    &$      13.95   _{\pm   5.75    }$&$    -1.11   _{\pm   0.60    }$&$    6.55    _{\pm   4.30    }$&$    -0.86   _{\pm   0.61    }$&$    8.99    _{\pm   4.81    }$&$    0.11    _{\pm   0.47    }$&$    4.07    _{\pm   0.80    }$&$    -0.56   _{\pm   0.20    }$\\
\ion{Ni}{I}     5082.339    &$      11.19   _{\pm   1.95    }$&$    -1.17   _{\pm   0.66    }$&$    7.60    _{\pm   1.74    }$&$    -1.12   _{\pm   0.61    }$&$    8.12    _{\pm   1.77    }$&$    -0.80   _{\pm   0.49    }$&$    6.77    _{\pm   0.43    }$&$    -0.83   _{\pm   0.19    }$\\
\ion{Fe}{I}     5367.467    &$      3.81    _{\pm   0.55    }$&$    -1.04   _{\pm   0.78    }$&$    6.31    _{\pm   0.58    }$&$    -1.02   _{\pm   0.78    }$&$    1.68    _{\pm   0.49    }$&$    -0.74   _{\pm   0.65    }$&$    3.82    _{\pm   0.16    }$&$    -0.81   _{\pm   0.27    }$\\
\ion{Fe}{I}     5373.709    &$      1.78    _{\pm   1.73    }$&$    -0.50   _{\pm   0.73    }$&$    3.63    _{\pm   1.64    }$&$    -0.69   _{\pm   0.70    }$&$    -3.02   _{\pm   1.54    }$&$    0.17    _{\pm   0.66    }$&$    3.74    _{\pm   0.38    }$&$    -0.68   _{\pm   0.25    }$\\
\ion{Fe}{I}     5383.369    &$      4.66    _{\pm   0.44    }$&$    -1.25   _{\pm   0.84    }$&$    7.61    _{\pm   0.46    }$&$    -1.42   _{\pm   0.88    }$&$    3.90    _{\pm   0.39    }$&$    -1.16   _{\pm   0.66    }$&$    6.65    _{\pm   0.14    }$&$    -1.03   _{\pm   0.28    }$\\
\ion{Ti}{II}        5418.751    &$      5.75    _{\pm   0.72    }$&$    -0.35   _{\pm   0.79    }$&$    5.75    _{\pm   0.65    }$&$    0.02    _{\pm   0.71    }$&$    5.59    _{\pm   0.61    }$&$    -0.24   _{\pm   0.59    }$&$    4.85    _{\pm   0.18    }$&$    0.13    _{\pm   0.25    }$\\
\ion{Fe}{I}     5576.089    &$      2.39    _{\pm   0.70    }$&$    -0.61   _{\pm   0.74    }$&$    4.67    _{\pm   0.68    }$&$    -0.94   _{\pm   0.84    }$&$    0.77    _{\pm   0.61    }$&$    -0.77   _{\pm   0.61    }$&$    3.45    _{\pm   0.16    }$&$    -0.89   _{\pm   0.30    }$\\
\ion{Fe}{I}     5862.353    &$      1.96    _{\pm   1.02    }$&$    0.09    _{\pm   0.84    }$&$    -0.18   _{\pm   1.11    }$&$    0.12    _{\pm   0.85    }$&$    -1.25   _{\pm   0.96    }$&$    0.01    _{\pm   0.66    }$&$    1.20    _{\pm   0.25    }$&$    -0.04   _{\pm   0.34    }$\\
\ion{Fe}{I}     6024.058    &$      5.70    _{\pm   0.77    }$&$    -1.11   _{\pm   0.89    }$&$    6.98    _{\pm   0.77    }$&$    -1.42   _{\pm   0.90    }$&$    4.58    _{\pm   0.67    }$&$    -1.26   _{\pm   0.71    }$&$    7.58    _{\pm   0.21    }$&$    -1.45   _{\pm   0.25    }$\\
\ion{Fe}{I}     6027.051    &$      4.68    _{\pm   1.59    }$&$    -0.52   _{\pm   0.78    }$&$    6.59    _{\pm   1.49    }$&$    -0.99   _{\pm   0.81    }$&$    3.71    _{\pm   1.31    }$&$    -0.56   _{\pm   0.61    }$&$    5.38    _{\pm   0.33    }$&$    -1.00   _{\pm   0.27    }$\\
\ion{Fe}{I}     6056.005    &$      3.02    _{\pm   1.53    }$&$    -0.07   _{\pm   0.82    }$&$    0.60    _{\pm   1.53    }$&$    -0.22   _{\pm   0.80    }$&$    2.33    _{\pm   1.40    }$&$    -0.44   _{\pm   0.65    }$&$    -0.43   _{\pm   0.35    }$&$    -0.38   _{\pm   0.30    }$\\
\ion{Si}{I}     6155.134    &$      -0.70   _{\pm   1.82    }$&$    0.31    _{\pm   0.61    }$&$    1.30    _{\pm   1.67    }$&$    -0.16   _{\pm   0.63    }$&$    -1.01   _{\pm   2.12    }$&$    0.29    _{\pm   0.47    }$&$    2.85    _{\pm   0.38    }$&$    -0.53   _{\pm   0.22    }$\\
\ion{Fe}{I}     6252.555    &$      5.19    _{\pm   0.68    }$&$    -0.93   _{\pm   0.93    }$&$    4.83    _{\pm   0.70    }$&$    -0.77   _{\pm   0.84    }$&$    5.99    _{\pm   0.61    }$&$    -1.30   _{\pm   0.70    }$&$    4.29    _{\pm   0.16    }$&$    -0.58   _{\pm   0.24    }$\\
\ion{Fe}{I}     6265.134    &$      3.63    _{\pm   1.10    }$&$    -0.63   _{\pm   0.83    }$&$    3.75    _{\pm   1.12    }$&$    -0.80   _{\pm   0.89    }$&$    1.11    _{\pm   0.92    }$&$    -0.70   _{\pm   0.67    }$&$    3.26    _{\pm   0.21    }$&$    -0.80   _{\pm   0.28    }$\\
\ion{Fe}{I}     6336.824    &$      8.99    _{\pm   1.01    }$&$    -1.45   _{\pm   0.89    }$&$    7.59    _{\pm   1.03    }$&$    -1.46   _{\pm   0.84    }$&$    6.62    _{\pm   0.88    }$&$    -1.75   _{\pm   0.65    }$&$    6.70    _{\pm   0.23    }$&$    -1.44   _{\pm   0.24    }$\\
  {\it average}          &$      4.34    _{\pm   1.87    }$&$                $&$ 4.23    _{\pm   1.63    }$&$                $&$ 2.61    _{\pm   1.52    }$&$                $&$ 3.69    _{\pm   0.33    }$&             \\
\hline \hline
Line (\AA)                   &         \multicolumn{2}{c}{$\zeta$~Gem}    &  \multicolumn{2}{c}{RZ~Vel}    &  \multicolumn{2}{c}{$\ell$~Car}       &   \multicolumn{2}{c}{RS~Pup}    \\
                             &  A$_{\gamma}$[\%] & V$_{\gamma}$[\kms]     &  A$_{\gamma}$[\%]&   V$_{\gamma}$[\kms]       &   A$_{\gamma}$[\%]&   V$_{\gamma}$[\kms]           &     A$_{\gamma}$[\%]&   V$_{\gamma}$[\kms]            \\
\hline
\ion{Fe}{I}     4683.560    &$      0.33    _{\pm   0.46    }$&$    -0.44   _{\pm   0.22    }$&$    -6.01   _{\pm   10.00   }$&$    0.24    _{\pm   0.99    }$&$    -1.78   _{\pm   0.49    }$&$    -0.32   _{\pm   0.32    }$&$    -5.97   _{\pm   3.18    }$&$    0.64    _{\pm   0.63    }$\\
\ion{Fe}{I}     4896.439    &$      -7.95   _{\pm   0.92    }$&$    0.96    _{\pm   0.24    }$&$    -5.89   _{\pm   10.86   }$&$    0.99    _{\pm   1.02    }$&$    2.10    _{\pm   0.86    }$&$    0.33    _{\pm   0.33    }$&$    -3.67   _{\pm   3.32    }$&$    1.35    _{\pm   0.75    }$\\
\ion{Fe}{I}     5054.643    &$      3.30    _{\pm   0.80    }$&$    -0.50   _{\pm   0.22    }$&$    3.15    _{\pm   7.02    }$&$    -0.22   _{\pm   1.01    }$&$    2.41    _{\pm   0.78    }$&$    -0.56   _{\pm   0.30    }$&$    0.22    _{\pm   2.75    }$&$    0.01    _{\pm   0.61    }$\\
\ion{Ni}{I}     5082.339    &$      6.51    _{\pm   0.44    }$&$    -0.62   _{\pm   0.20    }$&$    6.75    _{\pm   2.37    }$&$    -1.00   _{\pm   1.01    }$&$    4.42    _{\pm   0.64    }$&$    -0.76   _{\pm   0.27    }$&$    6.24    _{\pm   1.65    }$&$    -0.53   _{\pm   0.52    }$\\
\ion{Fe}{I}     5367.467    &$      3.95    _{\pm   0.19    }$&$    -0.43   _{\pm   0.30    }$&$    1.00    _{\pm   0.58    }$&$    -0.09   _{\pm   1.51    }$&$    -0.87   _{\pm   0.27    }$&$    0.13    _{\pm   0.40    }$&$    -0.03   _{\pm   0.39    }$&$    0.01    _{\pm   1.13    }$\\
\ion{Fe}{I}     5373.709    &$      3.56    _{\pm   0.41    }$&$    -0.59   _{\pm   0.28    }$&$    1.51    _{\pm   1.87    }$&$    0.03    _{\pm   1.30    }$&$    1.73    _{\pm   0.49    }$&$    -0.53   _{\pm   0.38    }$&$    -0.04   _{\pm   0.99    }$&$    0.15    _{\pm   0.94    }$\\
\ion{Fe}{I}     5383.369    &$      6.53    _{\pm   0.16    }$&$    -0.84   _{\pm   0.33    }$&$    4.01    _{\pm   0.47    }$&$    -0.79   _{\pm   1.32    }$&$    5.75    _{\pm   0.23    }$&$    -1.06   _{\pm   0.43    }$&$    8.66    _{\pm   0.31    }$&$    -0.71   _{\pm   1.07    }$\\
\ion{Ti}{II}        5418.751    &$      4.75    _{\pm   0.20    }$&$    0.45    _{\pm   0.28    }$&$    4.14    _{\pm   0.57    }$&$    0.17    _{\pm   1.43    }$&$    -0.62   _{\pm   0.31    }$&$    1.07    _{\pm   0.44    }$&$    4.65    _{\pm   0.45    }$&$    -0.52   _{\pm   0.96    }$\\
\ion{Fe}{I}     5576.089    &$      2.39    _{\pm   0.19    }$&$    -0.25   _{\pm   0.35    }$&$    -1.49   _{\pm   0.74    }$&$    -0.37   _{\pm   1.36    }$&$    -0.03   _{\pm   0.21    }$&$    0.04    _{\pm   0.50    }$&$    -3.67   _{\pm   0.47    }$&$    0.30    _{\pm   1.11    }$\\
\ion{Fe}{I}     5862.353    &$      -0.56   _{\pm   0.30    }$&$    0.35    _{\pm   0.37    }$&$    -2.89   _{\pm   1.21    }$&$    0.57    _{\pm   1.46    }$&$    -2.42   _{\pm   0.36    }$&$    0.43    _{\pm   0.49    }$&$    -6.09   _{\pm   0.72    }$&$    0.73    _{\pm   1.11    }$\\
\ion{Fe}{I}     6024.058    &$      6.80    _{\pm   0.24    }$&$    -1.01   _{\pm   0.32    }$&$    5.83    _{\pm   0.78    }$&$    -1.23   _{\pm   1.41    }$&$    7.54    _{\pm   0.31    }$&$    -1.24   _{\pm   0.44    }$&$    8.42    _{\pm   0.48    }$&$    -1.66   _{\pm   0.97    }$\\
\ion{Fe}{I}     6027.051    &$      4.19    _{\pm   0.37    }$&$    -0.43   _{\pm   0.32    }$&$    4.32    _{\pm   1.78    }$&$    -0.21   _{\pm   1.32    }$&$    2.73    _{\pm   0.42    }$&$    -0.38   _{\pm   0.45    }$&$    2.90    _{\pm   1.05    }$&$    -0.37   _{\pm   0.88    }$\\
\ion{Fe}{I}     6056.005    &$      -2.40   _{\pm   0.41    }$&$    0.15    _{\pm   0.34    }$&$    -6.75   _{\pm   1.79    }$&$    0.98    _{\pm   1.58    }$&$    -7.00   _{\pm   0.51    }$&$    0.68    _{\pm   0.45    }$&$    -8.53   _{\pm   0.85    }$&$    0.53    _{\pm   1.05    }$\\
\ion{Si}{I}     6155.134    &$      1.74    _{\pm   0.43    }$&$    -0.09   _{\pm   0.25    }$&$    -1.35   _{\pm   2.76    }$&$    0.71    _{\pm   1.21    }$&$    0.44    _{\pm   0.64    }$&$    0.08    _{\pm   0.36    }$&$    -1.49   _{\pm   1.14    }$&$    0.26    _{\pm   0.78    }$\\
\ion{Fe}{I}     6252.555    &$      3.22    _{\pm   0.19    }$&$    0.07    _{\pm   0.27    }$&$    0.81    _{\pm   0.75    }$&$    -0.16   _{\pm   1.40    }$&$    0.77    _{\pm   0.27    }$&$    0.22    _{\pm   0.34    }$&$    1.66    _{\pm   0.47    }$&$    -0.93   _{\pm   0.84    }$\\
\ion{Fe}{I}     6265.134    &$      2.20    _{\pm   0.24    }$&$    -0.38   _{\pm   0.35    }$&$    -0.13   _{\pm   1.30    }$&$    -0.21   _{\pm   1.44    }$&$    1.14    _{\pm   0.23    }$&$    -0.07   _{\pm   0.46    }$&$    0.23    _{\pm   0.60    }$&$    0.70    _{\pm   1.13    }$\\
\ion{Fe}{I}     6336.824    &$      4.70    _{\pm   0.27    }$&$    -0.87   _{\pm   0.26    }$&$    2.81    _{\pm   1.09    }$&$    -0.88   _{\pm   1.24    }$&$    3.22    _{\pm   0.35    }$&$    -0.59   _{\pm   0.35    }$&$    3.78    _{\pm   0.77    }$&$    -0.92   _{\pm   0.69    }$\\
    {\it average}          &$      2.54    _{\pm   0.37    }$&$                $&$ 0.58    _{\pm   2.70    }$&$                $&$ 1.15    _{\pm   0.43    }$&$                $&$ 0.43    _{\pm   1.15    }$&             \\

\hline \noalign{\smallskip}
\end{tabular}
\end{center}
\end{table*}

\end{document}